\documentclass[twocolumn,twocolappendix]{aastex7}

\usepackage{amsmath}	% Advanced maths commands

\newcommand{\shark}{\textsc{shark}}
\newcommand{\nth}{$^\mathrm{th}$}

\newcommand{\mstar}[1]{10^{#1}\ \mathrm{M}_\odot}
\defcitealias{robotham2011a}{R11}
\defcitealias{davies2018}{D18}

\submitjournal{ApJ}

\begin{document}

\title{Deep Extragalactic VIsible Legacy Survey (DEVILS): Galaxy group catalogue\\ for the D10-COSMOS field with 90\% spectroscopic redshift completeness}
\shorttitle{DEVILS: D10 galaxy group catalogue}

\author[0000-0001-5742-7927, gname=Mat\'ias, sname=Bravo]{Mat\'ias Bravo}
\affiliation{Department of Physics \& Astronomy, McMaster University, 1280 Main Street W, Hamilton, ON, L8S 4M1, Canada}
\email[show]{mabravo8@uc.cl}
\correspondingauthor{Mat\'ias Bravo}

\author[0000-0003-3085-0922, gname=Luke, sname=Davies]{Luke J. M. Davies}
\affiliation{International Centre for Radio Astronomy Research (ICRAR), M468, University of Western Australia,\\ 35 Stirling Hwy, Crawley, WA 6009, Australia.}
\email{luke.davies@icrar.org}

\author[0000-0003-0429-3579, gname=Aaron, sname=Robotham]{Aaron S. G. Robotham}
\affiliation{International Centre for Radio Astronomy Research (ICRAR), M468, University of Western Australia,\\ 35 Stirling Hwy, Crawley, WA 6009, Australia.}
\email{aaron.robotham@icrar.org}

\author[0000-0003-3021-8564, gname=Claudia, sname=Lagos]{Claudia del P. Lagos}
\affiliation{International Centre for Radio Astronomy Research (ICRAR), M468, University of Western Australia,\\ 35 Stirling Hwy, Crawley, WA 6009, Australia.}
\affiliation{ARC Centre of Excellence for All Sky Astrophysics in 3 Dimensions (ASTRO 3D).}
\affiliation{Cosmic Dawn Center (DAWN), Denmark}
\email{claudia.lagos@icrar.org}

\author[0000-0003-4169-9738, gname=Sabine, sname=Bellstedt]{Sabine Bellstedt}
\affiliation{International Centre for Radio Astronomy Research (ICRAR), M468, University of Western Australia,\\ 35 Stirling Hwy, Crawley, WA 6009, Australia.}
\affiliation{ARC Centre of Excellence for All Sky Astrophysics in 3 Dimensions (ASTRO 3D).}
\email{sabine.bellstedt@icrar.org}

\author[0000-0001-7516-4016, gname=Joss, sname=Bland-Hawthorn]{Joss Bland-Hawthorn}
\affiliation{Sydney Institute for Astronomy (SIfA), School of Physics, The University of Sydney, NSW 2006, Australia.}
\affiliation{ARC Centre of Excellence for All Sky Astrophysics in 3 Dimensions (ASTRO 3D).}
\email{jonathan.bland-hawthorn@sydney.edu.au}

\author[0000-0002-2949-2155, gname=Malgorzata, sname=Siudek]{Malgorzata Siudek}
\affiliation{Instituto de Astrof\'isica de Canarias, V\'ia L\'actea, 38205 La Laguna, Tenerife, Spain.}
\affiliation{Instituto de Astrof\'isica de Canarias (IAC); Departamento de Astrof\'isica, Universidad de La Laguna (ULL), 38200, La Laguna, Tenerife, Spain.}
\email{malgorzata.siudek@gmail.com}

\author[0000-0001-6263-0970, gname=Trystan, sname=Lambert]{Trystan S. Lambert}
\affiliation{International Centre for Radio Astronomy Research (ICRAR), M468, University of Western Australia,\\ 35 Stirling Hwy, Crawley, WA 6009, Australia.}
\email{trystan.lambert@icrar.org}

\author[0000-0002-4003-0904, gname=Chris, sname=Power]{Chris Power}
\affiliation{International Centre for Radio Astronomy Research (ICRAR), M468, University of Western Australia,\\ 35 Stirling Hwy, Crawley, WA 6009, Australia.}
\affiliation{ARC Centre of Excellence for All Sky Astrophysics in 3 Dimensions (ASTRO 3D).}
\email{chris.power@uwa.edu.au}

%\author{Simon Driver}
%\affiliation{International Centre for Radio Astronomy Research (ICRAR), M468, University of Western Australia,\\ 35 Stirling Hwy, Crawley, WA 6009, Australia.}
%\email{simon.driver@icrar.org}

\shortauthors{Bravo et al.}

\begin{abstract}
Large-scale galaxy redshift surveys conducted over the last couple of decades have proven crucial in deepening our understanding of structure growth in the Universe and galaxy evolution.
While there have been several such surveys, until now those that achieve the high completeness and precision necessary to probe the low-mass end of galaxy groups have been limited to relatively low redshifts ($z\lesssim0.3$), with surveys exploring the more distant Universe being constrained by small sample sizes and/or low redshift completeness.
The recent Deep Extragalactic VIsible Legacy Survey (DEVILS) aims to explore galaxy environment over the last $\sim6$ Gyr with a completeness level comparable to the most complete local Universe surveys ($>85\%$).
In this work, we present the galaxy group catalogue for the D10-COSMOS field from DEVILS, which achieves a redshift completeness of $90\%$ for galaxies with $Y<21.2$ mag.
We showcase the science potential by exploring the impact of environment on the fraction and power of active galactic nuclei (AGN), finding that satellites in galaxy groups show no evidence of altered AGN properties, while satellites in clusters exhibit increased AGN fractions but decreased AGN luminosities.
%225 words

\end{abstract}


%%%%%%%%%%%%%%%%%%%%%%%%%%%%%%%%%%%%%%%%%%%%%%%%%%
% Introduction
%%%%%%%%%%%%%%%%%%%%%%%%%%%%%%%%%%%%%%%%%%%%%%%%%%
\section{Introduction}

Galaxies form in density peaks of the matter distribution of the Universe, with peaks being themselves part of larger structures dominated by dark matter \citep[e.g.,][]{white1978}. 
To understand the properties of these structures and how they impact the galaxies that reside in them, a number of galaxy group catalogues have been constructed from the data of large-scale galaxy redshift surveys, using a variety of different algorithms \citep[usually referred to as group finders, e.g.,][]{yang2005,robotham2011a,tinker2022}.
Most group finders use one of two methods: a friend-of-friends algorithm that connect galaxies based on the distance between pairs \citep[e.g.,][]{huchra1982,eke2004,robotham2011a}, and halo-based finders that use our understanding of dark matter haloes to group galaxies \citep[e.g.,][]{yang2005,tinker2022}.
The resulting galaxy group catalogues have enabled the exploration of how galaxies are shaped by the environment in which they reside, improving our understanding of the physical processes shaping galaxy evolution \citep[e.g.,][]{knobel2014,robotham2014,davies2019b,oxland2024,foster2025}.
As galaxies also serve as tracers of larger dark matter structures, galaxy group catalogues are also crucial for the study of dark matter haloes that host these groups and their baryon content \citep[e.g.,][]{sifon2015,tojeiro2017,driver2022,dev2024}.

A spectroscopic redshift survey is needed to accurately measure galaxy environment down to the scale of small groups, as photometric redshift surveys lack the required precision in the redshift measurement, with errors in the $\sim3-10\%$ range, compared to the $<1\%$ required to detect groups \citep{mo2010book}.
Large surveys like the Sloan Digital Sky Survey \citep[SDSS; e.g.][]{york2000,abazajian2009}, the Two Degree Field Galaxy Redshift Survey \citep[2dFGRS]{colless2001} and the Galaxy and Mass Assembly survey \citep[GAMA;][]{driver2011,driver2022} have enabled the exploration of the environmental impact on galaxy evolution across a variety of halo sizes, though the limited redshift range covered by them restricts the exploration of the evolution of these effects with cosmic time to the very late Universe ($z\lesssim0.3$, i.e, the last $\sim3$ Gyr).
Other surveys have traded sky coverage for depth, like the redshift Cosmic Evolution Survey \citep[zCOSMOS;][]{lilly2007}, Very Large Telescope VIsible Multi-Object Spectrograph Deep Survey \citep[VLT/VIMOS Deep Survey or VDSS;][]{lefevre2013}, and VIMOS Ultra-Deep Survey \citep[VUDS;][]{lefevre2015}, enabling the exploration of higher redshifts.

A common limitation of these pencil-beam surveys is the comparatively low redshift completeness ($\lesssim50\%$), i.e., how many of the targeted galaxies have a secure redshift measured, compared to surveys like SDSS and GAMA ($\sim70\%$ and $95\%$, respectively).
A high completeness is crucial to accurately explore the properties of galaxy groups, as without this, the properties of groups can be subjected to large errors from poor statistics, and surveys can even completely miss groups due to a lack of measurements of enough members.
The recent Deep Extragalactic VIsible Legacy Survey \citep[DEVILS;][]{davies2018} aims to improve the quality of environmental measurements at higher redshift, combining the pencil-beam design of other high-$z$ spectroscopic surveys with the high redshift completeness of existing low-$z$ surveys, providing observations of comparable quality to those of GAMA and serving as a pathfinder for the deep component of the upcoming Wide Area Vista Extragalactic Survey \citep[WAVES;][]{driver2019}.

In this work, we present the first version of the galaxy group catalogue from DEVILS.
We describe the DEVILS-D10 galaxy catalogue and DEVILS-like synthetic lightcones we use in this work in Section \ref{S2:base_cats}, provide an overview of the group finder first used for the GAMA survey \citep[hereafter R11]{robotham2011a} in Section \ref{S3:groupfinder}, outline our calibration of the \citetalias{robotham2011a} group finder in Section \ref{S4:calibration}, present the resulting DEVILS-D10 galaxy group catalogue in Section \ref{S5:group_cat}, briefly explore the environmental impact on galaxies in Section \ref{S6:agn_env}, and close with our conclusions in Section \ref{S5:conclusion}.

%%%%%%%%%%%%%%%%%%%%%%%%%%%%%%%%%%%%%%%%%%%%%%%%%%
% DEVILS & SHARK
%%%%%%%%%%%%%%%%%%%%%%%%%%%%%%%%%%%%%%%%%%%%%%%%%%
\section{Galaxy and calibration catalogues}\label{S2:base_cats}

In this work, we make use of both observed and synthetic galaxy catalogues produced by the DEVILS team, with the latter tailored to mimic the DEVILS footprint and selection for calibration of the group finder parameters.
Here we provide a brief overview of the data we use from our D10 observations (Section \ref{S2.1:DEVILS}) and our synthetic DEVILS-like lightcones (Section \ref{S2.2:LCs}). 

%%%%%%%%%%%%%%%%%%%%%%%%%%%%%%%%%%%%%%%%%%%%%%%%%%
\subsection{DEVILS-D10}\label{S2.1:DEVILS}

\begin{figure*}
    \centering
    \includegraphics[width=\linewidth]{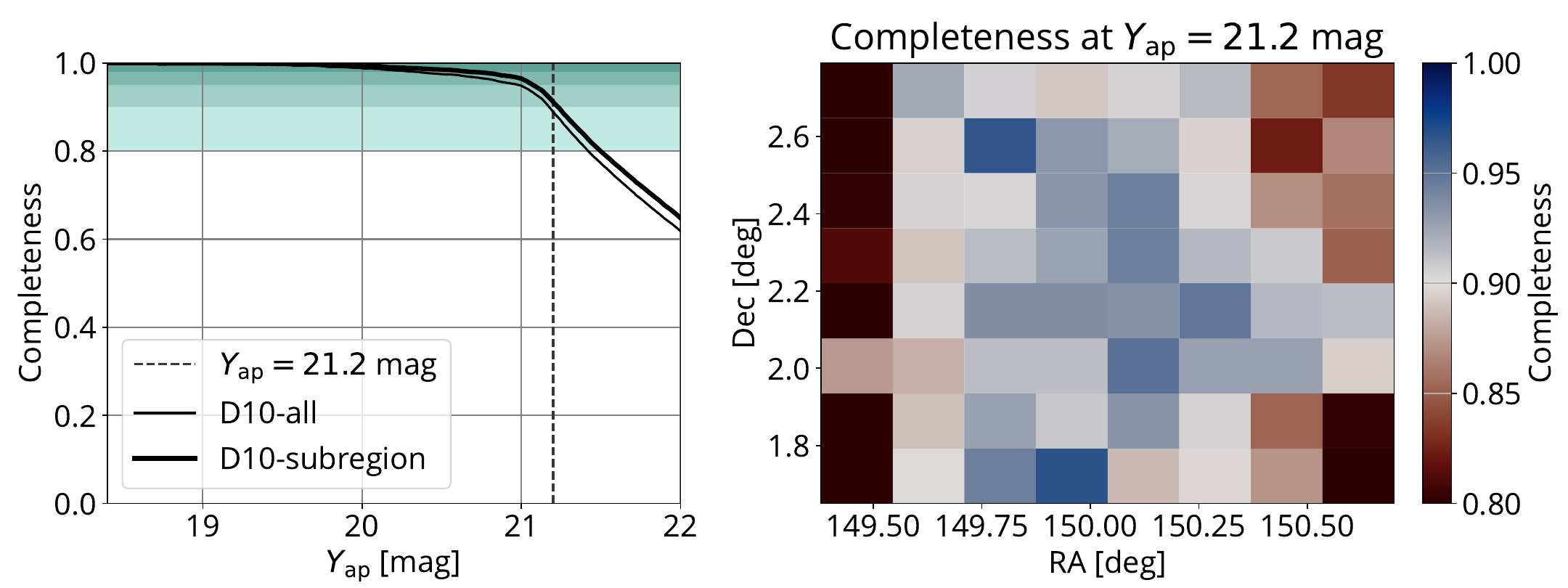}
    \caption{\textbf{Left:} The redshift completeness in DEVILS-D10 as a function of $Y$-band magnitude. Solid lines indicate the completeness for the whole D10 region (thin line) and the central subregion (solid line, defined as $149.545^\circ\leq\alpha\leq150.535^\circ$). The dashed vertical line indicates the survey target magnitude of $Y=21.2$ mag. The horizontal shaded regions indicate, from light to dark, the 80/90/90/98\% completeness region. \textbf{Right:} The redshift completeness in DEVILS-D10 at the survey target limit, $Y=21.2$ mag, as a function of right ascension and declination. Bins are coloured to highlight the difference from 90\% completeness, with blue (red) shades indicating higher (lower) completeness.}
    \label{fig:D10_comp}
\end{figure*}

The \textit{Deep Extragalactic VIsible Legacy Survey} (DEVILS; \citetalias{davies2018}) is an intermediate-redshift ($z_\mathrm{median}\sim0.6$) spectroscopic survey, with the main aim of producing high-completeness redshift catalogues for the study of the evolution of halo masses, merger rates, and environmental effects on galaxy evolution.
It was carried out using the AAOmega spectrograph \citep{saunders2004,sharp2006}, fed with fibres positioned with the Two-degree Field \citep[2dF;][]{lewis2002}, on the $3.9\ m$ Anglo-Australian Telescope (AAT) in Siding Spring Observatory.
The survey targeted $\sim60,000$ galaxies in three well-studied fields: Cosmic Evolution Survey \citep[COSMOS;][]{scoville2007}, Extended Chandra Deep Field South \citep[ECDFS;][]{lehmer2005} and X-ray Multi-mirror Mission Large-Scale Structure \citep[XMM-LSS;][]{pierre2002}.
These fields were chosen due to the wide variety of existing and planned observations across the electromagnetic spectrum, where the addition of high-quality environmental metrics would maximise the science output of the survey.
In this work, we focus on the part of the COSMOS field covered by DEVILS, which we will refer to as D10 from here onwards \citep[see table 1 in][for the detailed description of D10]{davies2018}, which will complement the existing variety of galaxy properties derived in previous works \citep{davies2021,hashemizadeh2021,thorne2021,thorne2022,cook2025}.

The specific aims chosen for DEVILS were measuring the evolution of $M^*_{z=0}$-like galaxies up to $z\sim1$, identifying all close pairs of $M^*_{z=0}$-like galaxies out to $z\sim0.8$, and measuring the mass of groups and clusters ($M_\mathrm{halo}\gtrsim\mstar{13}$) out to $z\sim0.7$.
From the tests carried out in \citetalias{davies2018}, a selection (apparent) magnitude of $Y<21.2$ mag was chosen for DEVILS, using photometry obtained with the Visible and Infrared Survey Telescope for Astronomy \citep[VISTA;][]{sutherland2015} for the UltraVISTA \citep[]{mccracken2012} and the VISTA Deep Extragalactic Observations \citep[VIDEO;][]{jarvis2013} surveys.
The target redshift completeness for DEVILS was $>95\%$, aiming for a comparable completeness to the GAMA survey \citep{driver2011,driver2022}, with a minimum required completeness of $85\%$.

Figure \ref{fig:D10_comp} shows the completeness for D10 after the end of the observation campaign in 2022, where we fall just short of a 90\% completeness at our target selection of $Y=21.2$ mag.
This is driven by two narrow strips at the western and eastern edges of the field, which were de-prioritised in the late stages of the survey campaign to account for time losses due to both weather and technical issues.
We define a sub-region of D10 to ignore these two strips by narrowing the right ascension range to $149.545^\circ\leq\alpha\leq150.535^\circ$ (the inner 6 columns of bins in the right panel of Figure \ref{fig:D10_comp}), where we measure a completeness slightly above 90\% at $Y=21.2$.
Irrespective of whether we include these two low-completeness strips or not, the overall completeness in D10 is well above our target of $85\%$.
While this completeness level falls short of the standard set by GAMA \citep[originally 98\% for $r<19.8$ mag, recently revised down to 95\% for $r<19.65$ mag;][]{driver2011,driver2022}, we choose to retain the magnitude limit of $Y=21.2$ for the D10 group catalogue as a balance between number of selected galaxies and the required completeness for our science goals, assuming a global completeness of 90\% for the field.
This magnitude limit leads to an input catalogue containing $15,183$ galaxies.

%%%%%%%%%%%%%%%%%%%%%%%%%%%%%%%%%%%%%%%%%%%%%%%%%%
\subsection{DEVILS-like synthetic lightcones}\label{S2.2:LCs}

Like most group finders, the \citetalias{robotham2011a} implementation includes a set of free parameters that need to be fine-tuned, which needs to be done on a per-survey basis.
This calibration process usually involves the use of synthetic lightcones, for which the association of each galaxy to DM haloes is known and used as ground truth for the calibration.
To this end, we generated our DEVILS-like lightcones using the combination of the \shark\ semi-analytic model \citep{lagos2018,lagos2024}, \textsc{stingray} lightcone builder \citep{stingray_043}, and \textsc{prospect} SED-generating tool \citep{robotham2020}.
The complete description of the process was presented in \citet{chauhan2019}, \citet{lagos2019}, and \citet{bravo2020}, so we present a brief summary here, highlighting differences with our previous methodology.

We start with the synthetic galaxy catalogues we generated with \shark, produced by running the latest version \citep[v2.0;][]{lagos2024} on a new version of the medi-SURFS run presented in \citet{elahi2018a}, which has a cosmological volume of $(210\ \mathrm{Mpc}/h)^3$ and $1536^3$ dark matter particles with a mass of $2.21\times10^8$.
This new version was run using the SWIFT code \citep{schaller2024}, and halo and subhalo catalogues and merger trees were constructed using HBT-HERONS \citep{forouhar2025,chandro2025}.
\citet{chandro2025} demonstrated that these catalogues were better suited for semi-analytic modelling than the previous generation of catalogues, which were used in the original SURFS suite, as they more cleanly track the merger trees of halos throughout the whole dynamic range included in these simulations.
In addition, we also incorporated the modifications to Shark introduced by \citet{chandro2025}, which mitigate the remaining issues in merger trees (including satellite/central subhalo tag swapping and some various missed links between halos at different snapshots).
However, we stress that those happen very rarely in our current N-body suite thanks to the improved performance of HBT-HERONS.

We used these outputs to generate 32 lightcones with \textsc{stingray}, each with the same sky footprint of DEVILS (i.e., all three fields), which roughly equals the total volume of $\sim1$ Gpc$^3$ used in \citetalias{robotham2011a} in the calibration for GAMA.
While not formally independent due to the use of a single simulation box, the pencil-beam design of DEVILS means that, combined with the unique random orientations and rotations applied to each lightcone, in practice, they are close to being independent.
Lastly, we used \textsc{prospect} \citep[through the wrapper \textsc{viperfish}][]{viperfish_053} to generate broad-band photometry for each galaxy in the lightcones, generating the full set of 28 bands covering from far-ultraviolet to far-infrared \citep[for a complete description of the bands see][]{davies2021}.

\begin{figure}
    \centering
    \includegraphics[width=\linewidth]{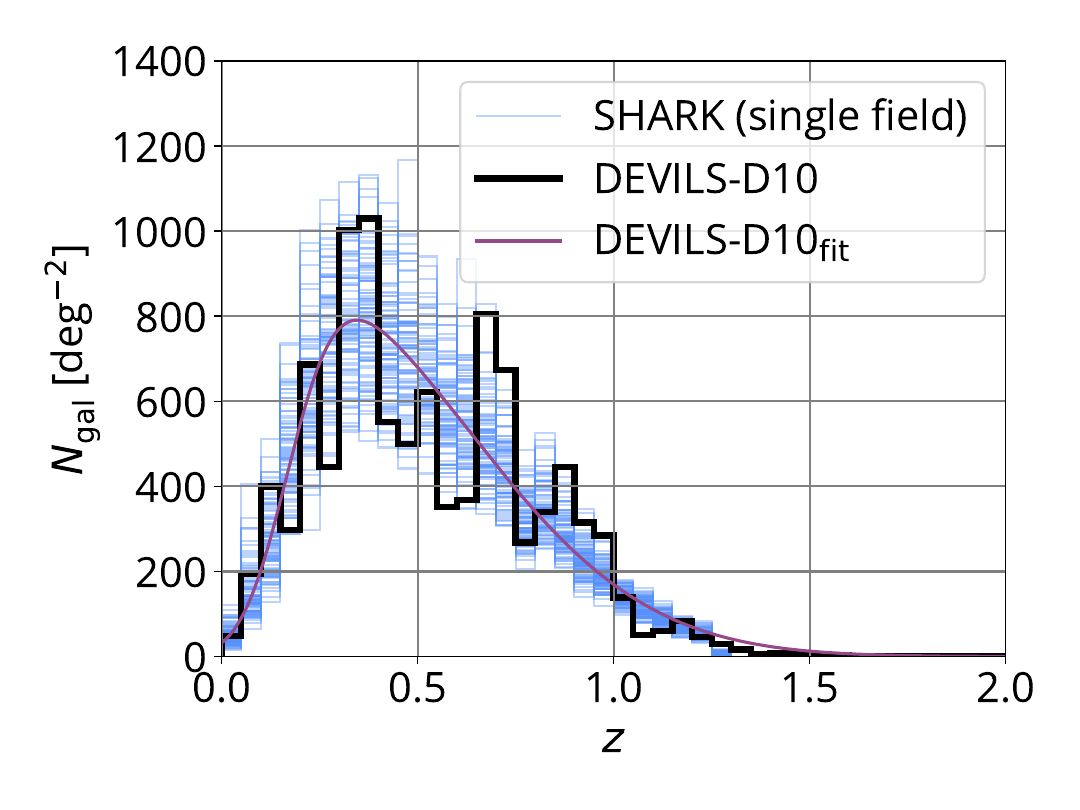}
    \caption{The galaxy redshift distribution in the DEVILS D10 field compared to the synthetic lightcones made with \textsc{shark}+\textsc{stingray}+\textsc{prospect} used for the calibration of the group finder.
    The solid black histogram shows the true D10 redshift distribution, the purple curve the smooth fit to the D10 distribution used to abundance-match the synthetic lightcones, and the faint blue histograms the individual fields from each synthetic lightcone after abundance-matching.}
    \label{fig:SHARK_DEVILS}
\end{figure}

As mentioned before, \shark\ reproduces a large number of observables well, but it is critical to ensure that the number density of galaxies in our synthetic lightcones matches that of D10 for the group finder calibration.
Therefore, we adopt a modification of the abundance matching we implemented in \citet{bravo2020}, with two key differences:
\begin{enumerate}
    \item We lack an equivalent to the random catalogue available for GAMA \citep{farrow2015} to use for the abundance matching, which we replace with a simpler model of the redshift distribution in D10 by fitting it with a skewed Gaussian using the statistics module in \textsc{scipy}.
    We show the true redshift distribution and our fit in Figure \ref{fig:SHARK_DEVILS}. 
    \item Since our simplified modelling only provides the number density at a given redshift, without the additional apparent magnitudes included in the GAMA random catalogue, we perform the abundance matching by shifting the $Y$-band apparent magnitudes of all galaxies in a given redshift bin by the same amount instead of applying the individual shifts we used in \citet{bravo2020}.
\end{enumerate}
We refer the reader to section 2.2 of \citet{bravo2020} for further details.   
We also show in Figure \ref{fig:SHARK_DEVILS} the abundance-matched redshift distributions from each field from our 32 synthetic lightcones, which qualitatively match the amplitude of the fluctuations seen in D10.

%%%%%%%%%%%%%%%%%%%%%%%%%%%%%%%%%%%%%%%%%%%%%%%%%%
% Group finder
%%%%%%%%%%%%%%%%%%%%%%%%%%%%%%%%%%%%%%%%%%%%%%%%%%
\section{The R11 Group Finder}\label{S3:groupfinder}

Following the lead of GAMA, we use a lightly updated version\footnote{The updates we have made to the group finder do not alter the logic of the algorithm, with the changes focusing on improving the implementation. These changes are discussed in more depth in Lambert et al. (in preparation).} of the \citetalias{robotham2011a} group finder.
The \citetalias{robotham2011a} group finder uses a friend-of-friends (FoF) algorithm, where pairs of galaxies that are closer than some predefined limit are linked together, and the collection of all interconnected galaxies is defined as a group.
To account for the effect of peculiar velocities in measuring distances with redshift, the \citetalias{robotham2011a} group finder defines separate maximum linking lengths for the directions perpendicular and parallel to the line of sight.
For links in each direction to be established, the following conditions must be satisfied:

\begin{align}
    \tan(\theta_{i,j})\left(\frac{D_{\mathrm{com},i}+D_{\mathrm{com},j}}{2}\right) &\leqslant \frac{b\left(D_{\mathrm{lim},i}+D_{\mathrm{lim},j}\right)}{2},\label{eq:01}\\
    \left|D_{\mathrm{com},i}-D_{\mathrm{com},j}\right| &\leqslant \frac{br\left(D_{\mathrm{lim},i}+D_{\mathrm{lim},j}\right)}{2},\label{eq:02}
\end{align}

\noindent where $\theta_{i,j}$ is the angle in the sky between galaxies $i$ and $j$, $D_{\mathrm{com},i}$ the radial distance to galaxy $i$ in comoving units, $D_{\mathrm{lim},i}$ the mean separation between galaxies at the position of galaxy $i$, and $b$ and $r$ are scaling factors for the links perpendicular and parallel to the line of sight, respectively.
$D_{\mathrm{lim},i}$ was originally defined in \citetalias{robotham2011a} as:

\begin{equation}
    D_{\mathrm{lim},i} = \left(\frac{\phi(M_{\mathrm{lim},i})}{\phi(M_{gal,i})}\right)^{\frac{\nu}{3}}\left(\int^{M_{\mathrm{lim},i}}_{-\infty}\phi(M)dM\right)^{-\frac{1}{3}},\label{eq:03}
\end{equation}

\noindent where $M_{\mathrm{lim},i}$ is the absolute magnitude limit of the survey at the position of galaxy $i$, $\phi(M)$ is the luminosity function (LF) of the survey and $M_{gal,i}$ is the absolute magnitude of galaxy $i$.
The term before the integral allows for larger linking lengths for intrinsically brighter galaxies, for which we expect more significant associations, where $\frac{\nu}{3}$ is the power law slope controlling the strength of this correction.
Since currently there is no measured LF for DEVILS, in this work, we simplify this calculation by adopting $\nu=0$ (i.e., no luminosity dependence) and use our redshift distribution fit described in Section \ref{S2.2:LCs} to replace the integration of the LF, effectively replacing Equation \ref{eq:03} with:

\begin{equation}
    D_{\mathrm{lim},i} = n(z^{}_i)^{-\frac{1}{3}}, \label{eq:03rev}
\end{equation}
\noindent where $n(z^{}_i)$ is the galaxy number density at the redshift of galaxy $i$.

While $b$ and $r$ can be treated as simple free parameters, \citetalias{robotham2011a} define both as a function of the local density contrast:

\begin{align}
    b(\vec{r},m_{\mathrm{lim}}) &= \frac{b_0}{c(\vec{r})^{\frac{1}{3}}}\left(\frac{1}{\Delta}\frac{\rho_{emp}(\vec{r},m_{\mathrm{lim}})}{\bar{\rho}(\vec{r},m_{\mathrm{lim}})}\right)^{E_b},\label{eq:04} \\
    r(\vec{r},m_{\mathrm{lim}}) &= r_0\left(\frac{1}{\Delta}\frac{\rho_{emp}(\vec{r},m_{\mathrm{lim}})}{\bar{\rho}(\vec{r},m_{\mathrm{lim}})}\right)^{E_R},\label{eq:05}
\end{align}

\noindent where $b_0$ $r_0$ are scaling constants, $\bar{\rho}$ is the local density implied by the magnitude limit of the survey at the position $\vec{r}$, $\rho_{emp}$ empirically estimated local density, $m_{\mathrm{lim}}$ is the apparent magnitude limit and $c(\vec{r})$ is the survey completeness near position $\vec{r}$.
Since we assume a completeness of 90\% across DEVILS-D10, $c(\vec{r})=0.9$ for all our selected galaxies.
$\rho_{emp}$ is estimated in a cylinder along the line of sight with radius and height $r_\Delta$ and $l_\Delta$, respectively, $\Delta$ is the ratio between local and global densities, and $E_b$ and $E_r$ are the slopes of each power law.
Similar to Equation \ref{eq:03}, we replace the use of the survey LF with our redshift distribution fit to calculate $\rho_{emp}$.
This leads to a total of seven free parameters that need to be optimised: $b_0$, $r_0$, $\Delta$, $r_\Delta$, $l_\Delta$, $E_b$, and $E_r$.

The \citetalias{robotham2011a} group finder also provides a number of additional group properties, calculated after the FoF group finding has been completed.
Unlike the FoF component, we make no changes to the derivation of group properties as presented in \citetalias{robotham2011a}.
In brief, these properties are:
\begin{itemize}
    \item Group velocity dispersions are calculated using the GAPPER estimator introduced in \citet{beers1990}, corrected for velocity measurement errors.
    \item Projected centres of the groups are identified with three different methods (centre of luminosity, BCG, and iterative refinement), and radial centres are identified with two different methods (median and iterative refinement).
    \item Projected sizes are defined as the minimum distance from the iteratively-refined centres that contains 50/68/100\% of the satellites of the group.
    \item Halo masses are estimated using the corrected velocity dispersion and the 50\% radii, including two alternative corrections to reduce the median bias between the true and measured halo mass in the synthetic lightcones.
\end{itemize} 
We refer the reader to sections 4.1, 4.2, and 4.3 for further details on the calculation of the velocity dispersions, centre and sizes, and halo masses, respectively.

We make the first public release of the code for the \citetalias{robotham2011a} group finder with this work.
We have archived static versions of the code in its last iteration used for the GAMA survey \citep{fof_101} and for the version used in this work \citep{fof_122}, and have made public the GitHub repository hosting the code\footnote{\url{https://github.com/ICRAR/FoF}}.
We note that we do not currently plan to continue developing this code, with these two versions being released for legacy purposes.
\textsc{nessie}\footnote{\url{https://github.com/TrystanScottLambert/Nessie}} (Lambert et al., submitted) is the successor for the \citetalias{robotham2011a} group finder in active development.

%%%%%%%%%%%%%%%%%%%%%%%%%%%%%%%%%%%%%%%%%%%%%%%%%%
% FoF calibration
%%%%%%%%%%%%%%%%%%%%%%%%%%%%%%%%%%%%%%%%%%%%%%%%%%
\section{Group finder calibration}\label{S4:calibration}

Since having a significant number of spurious groups in the final catalogue would negatively affect most of the science cases using groups, the priority of this optimisation is not to retrieve perfectly all the groups from the simulation, but to have both original (simulation) and found (group finder) groups be an accurate representation of each other.
To this end, we use the figure of merit (FoM) used in \citetalias{robotham2011a} for the calibration of the free parameters.
This FoM is constructed from two measurements of the quality of the group recovery: the fraction of bijectively-matched groups and the purity of the groups.

\begin{deluxetable}{LCCc}
    \tablehead{
        \colhead{Parameter} & \colhead{Boundaries} & \colhead{Value} & \colhead{Method} \\
    }
    \tablecaption{
        Optimisation boundaries and adopted values for the free parameters in the \citetalias{robotham2011a} group finder.\label{tab:calparam}
    }
    \startdata
        \log_{10}(b_0) & [-2.00,-0.50] & -1.25477   & Optimised \\
        r_0            & [10.0,65.0]   & 52.3511    & Optimised \\
        E_b            & [-0.40,0.40]  & -0.0751399 & Optimised \\
        E_r            & [-0.40,0.40]  & 0.116167   & Optimised \\
        \nu            &       -       & 0.0        & Fixed \\
        \Delta         &       -       & 9.0        & Fixed \\
        r_\Delta       &       -       & 1.5        & Fixed \\
        l_\Delta       &       -       & 12         & Fixed 
    \enddata
\end{deluxetable}

\begin{figure*}
    \centering
    \includegraphics[width=\linewidth]{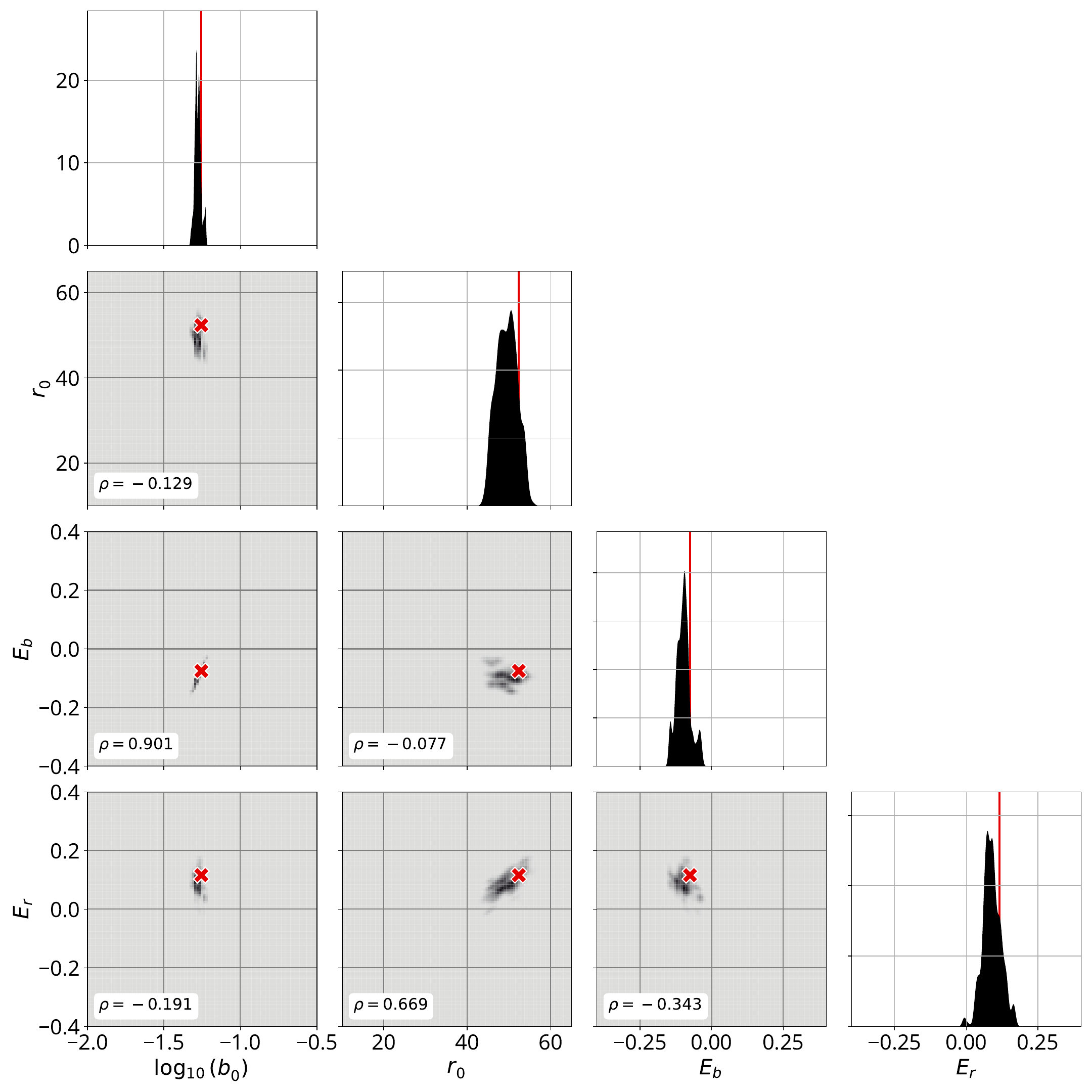}
    \caption{Distributions of the posterior chain for the last MCMC optimisation step of the four main free parameters ($b_0$, $r_0$, $E_b$, and $E_r$).
    The diagonal panels show the individual distribution of each parameter, smoothed with a 1D Gaussian kernel density estimator with the black curves, and the values for the highest $S_\mathrm{tot}$ with red lines.
    The non-diagonal panels show combined distributions of each parameter pair smoothed with a 2D Gaussian kernel density estimator with the greyscale histogram, and the values for the highest $S_\mathrm{tot}$ with red markers.
    The axes span the complete range of values explored in the optimisation process.
    The inset text on the bottom left of the non-diagonal panels indicates the Spearman Rank Coefficient ($\rho$) for each parameter pair.
    $p$-values are not shown as they are small in all cases ($p\lesssim10^{-4}$).}
    \label{fig:DEVILS_DR1_ParamCal}
\end{figure*}

A pair of groups, one from the simulation halo finding and another from the \citetalias{robotham2011a} group finder, is defined as bijectively matched when the shared galaxies between them are at least $50\%$ of the total number of members of each:
\begin{equation*}
    E^{}_{\{\mathrm{LC,GF}\}} = \frac{N_{\mathrm{group,bij}}}{N_{\mathrm{group,\{LC,GF\}}}},
\end{equation*}
\noindent where $N_{\mathrm{group,bij}}$ is the number of bijectively-matched groups, $N_{\mathrm{group,\{LC,GF\}}}$ is the total number of groups using either the true haloes in the lightcone or those found by the \citetalias{robotham2011a} group finder, and $E^{}_{\{\mathrm{LC,GF}\}}$ is the resulting bijective fraction for both true and \citetalias{robotham2011a} group catalogues.
We note that $E^{}_{\{\mathrm{LC,GF}\}}$ have values between 0 (no bijective groups found) and 1 (all groups are bijective).
The group purity is quantified as:
\begin{align*}
    &P^{}_{\{\mathrm{LC,GF}\}}[i,j] = \frac{N_{\mathrm{gal,shared}}[i,j]}{N_{\mathrm{gal,\{LC,GF\}}}[i]} \frac{N_{\mathrm{gal,shared}}[i,j]}{N_{\mathrm{gal,\{GF,LC\}}}[j]}, \\
    &P^{}_{\{\mathrm{LC,GF}\}}[i] = \max\left( \left\{ P^{}_{\{\mathrm{LC,GF}\}}[i,j] \right\}_{j=1..N_{\mathrm{group,\{GF,LC\}}}} \right), \\
    &Q^{}_{\{\mathrm{LC,GF}\}} = \frac{\sum^{N_{\mathrm{g},\{\mathrm{LC,GF}\}}}_{i=1}P^{}_{\{\mathrm{LC,GF}\}}[i]N_{\mathrm{gal,\{LC,GF\}}}[i]}{\sum\ N_{\mathrm{gal,\{LC,GF\}}}},
\end{align*}
\noindent where $N_{\mathrm{gal,shared}}[i,j]$ is the number of galaxies shared between the $i$-th group in the lightcone (\citetalias{robotham2011a} group finder) and the $j$-th group in the \citetalias{robotham2011a} group finder (lightcone), $N_{\mathrm{gal,\{LC,GF\}}}[i]$ is the number of galaxies in the $i$-th group in the lightcone (\citetalias{robotham2011a} group finder), and $Q^{}_{\{\mathrm{LC,GF}\}}$ is the resulting group purity for either lightcone or \citetalias{robotham2011a} group finder.
We note that, as with $E^{}_{\{\mathrm{LC,GF}\}}$, the values for $Q^{}_{\{\mathrm{LC,GF}\}}$ can range from 0 (no groups in either catalogue share galaxies with those in the other) to 1 (all groups are recovered with all their members and no interlopers).
The FoM is then defined using these four values as:
\begin{align*}
    E^{}_\mathrm{tot} &= E^{}_\mathrm{LC}E^{}_\mathrm{GF},\\
    Q_\mathrm{tot} &= Q^{}_\mathrm{LC}Q^{}_\mathrm{GF},\\
    S_\mathrm{tot}&=E^{}_\mathrm{tot}Q^{}_\mathrm{tot}.
\end{align*}

\begin{figure*}
    \centering
    \includegraphics[width=\linewidth]{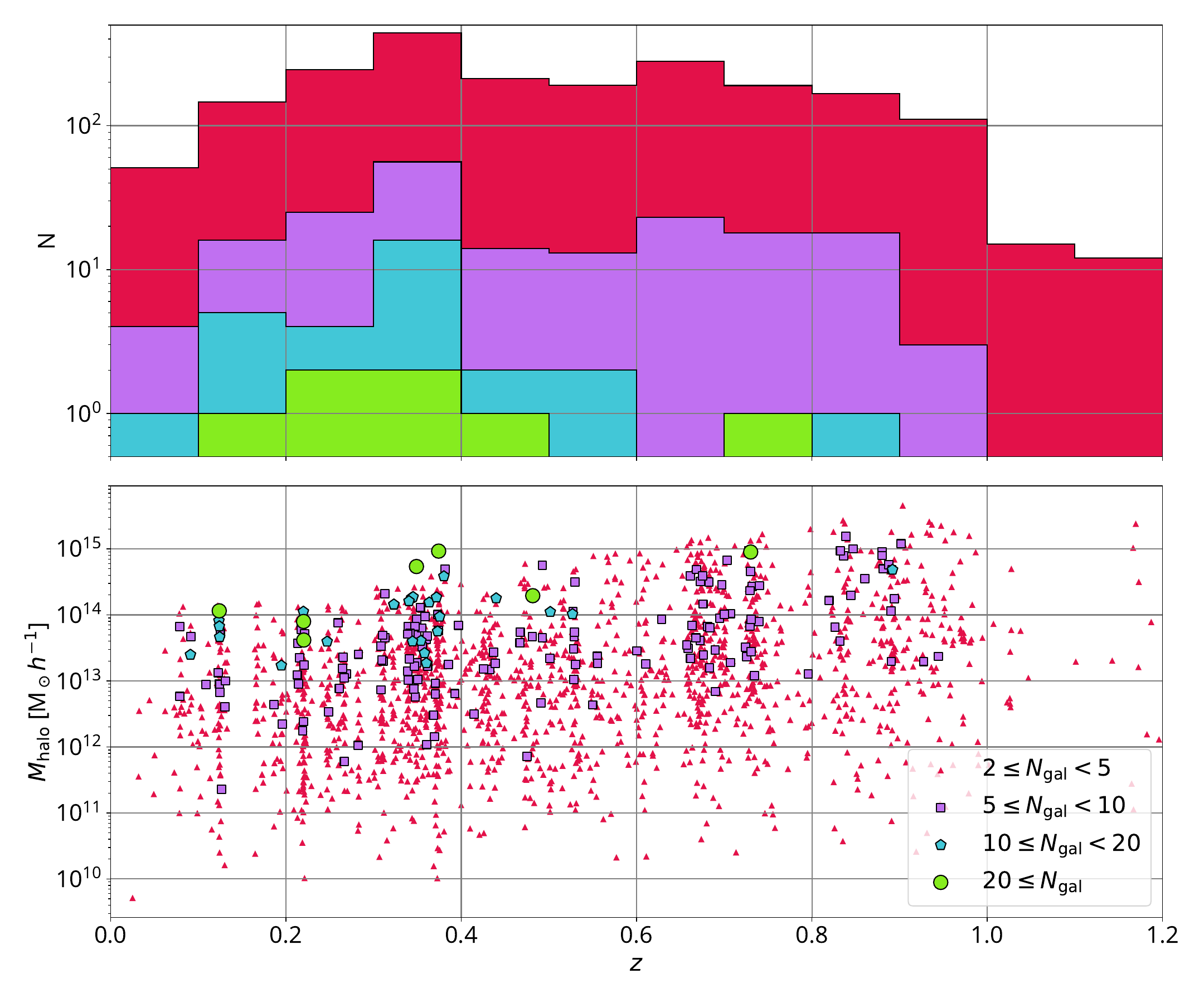}
    \caption{Galaxy groups in the DEVILS D10 group catalogue.
    The top panel shows stacked histograms of the redshift distribution of the galaxy groups in the DEVILS D10 field, and the bottom panel show the group halo masses as a function of redshift.
    Galaxy groups are displayed according to the number of member galaxies, groups with ($2\leq N_\mathrm{gal}<5$) in magenta (small triangles in the bottom panel), groups with $5\leq N_\mathrm{gal}<10$ in purple (medium-small squares), groups with $10\leq N_\mathrm{gal}<20$ in teal (medium-large pentagons), and groups with $20\leq N_\mathrm{gal}$ in green (large circles).
    The halo masses in the bottom panel are the proxy for dynamical group mass, corrected for median biases (see \citetalias{robotham2011a} for further details).}
    \label{fig:D10_RAz}
\end{figure*}

We use the \textsc{highlander} \citep{highlander_0210} to find the values of the free parameters that maximise $S_\mathrm{tot}$, which uses a combination of covariance matrix adaptation evolution strategy (CMAES) and a component-wise hit and run Metropolis (CHARM) implementation of Markov chain Monte Carlo (MCMC).
As in \citetalias{robotham2011a}, we use only groups with at least five members ($N_\mathrm{gal}\geq5$) for the calibration, and we also restrict the calibration to groups in the redshift range spanned by the 10\nth\ and 90\nth\ percentiles of the D10 galaxy redshift distribution ($0.213<z<0.910$).
Early experimentation with the calibration highlighted that a number of degeneracies exist between some of the remaining free parameters, in particular between the two main parameters ($b_0$ and $r_0$) and the three density contrast parameters ($\Delta$, $\Delta_r$, and $\Delta_l$).
The posteriors of the latter were comparatively wide and consistent with the fixed values adopted in \citetalias{robotham2011a}, so we chose to use the same values for these three parameters (see Table \ref{tab:calparam}), and focus the calibration on the main four parameters: $b_0$, $r_0$, $E_b$, and $E_r$.
The adopted boundaries for the optimisation of each parameter are included in Table \ref{tab:calparam}.

The final calibration with \textsc{highlander} was performed in a similar manner to how it has been used for SED fitting in \citet{thorne2021,thorne2022}, using two rounds of CMAES-CHARM/MCMC optimisation, with all using 250 iterations, save for the last MCMC round that uses 2500 iterations (to better sample the posteriors).
We show the posterior distributions for the four parameters from this last MCMC round in Figure \ref{fig:DEVILS_DR1_ParamCal}, which shows that all four have converged to a narrow range of values.
We note that we find clear correlations between two sets of parameters ($b_0$-$E_b$ and $r_0$-$E_r$), but we decide to keep these values since the posteriors of both $E_b$ and $E_r$ exclude the default choice made in GAMA ($E_b=E_r=0$).
The highest value of the FoM we achieve is $S_\mathrm{tot}=0.1697$, with $E^{}_\mathrm{LC}$ and $Q^{}_\mathrm{LC}$ being systematically higher than $E^{}_\mathrm{GF}$ and $Q^{}_\mathrm{GF}$ ($\sim0.75$ and $\sim0.55$, respectively), indicating that this calibration mildly prioritises completeness over purity.
We present the values corresponding to the highest FoM in Table \ref{tab:calparam}.

%%%%%%%%%%%%%%%%%%%%%%%%%%%%%%%%%%%%%%%%%%%%%%%%%%
% Group catalogue
%%%%%%%%%%%%%%%%%%%%%%%%%%%%%%%%%%%%%%%%%%%%%%%%%%
\section{DEVILS-D10 Group Catalogue}\label{S5:group_cat}

Using the parameter values from Table \ref{tab:calparam} for the \citetalias{robotham2011a} group finder, we find that, of the $15,183$ selected galaxies from DEVILS-D10, $9,262$ are isolated galaxies, with the remaining $5,921$ galaxies being distributed across $2,060$ groups
Of these groups, $1324$ groups are pairs, $546$ have 3-4 members, and $190$ have at least 5 members.
Of the groups with $N_\mathrm{gal}\geq5$, the closest one is located at $z=0.0787$ and the farthest one at $z=0.9445$, based on the median redshift of their members.
There are 7 groups with $N_\mathrm{gal}\geq20$, spanning the $0.1\lesssim z\lesssim0.7$ range with halo masses in the $10^{13.5}\lesssim M_\mathrm{halo}/(\mathrm{M}_\odot h^{-1})\lesssim10^{15}$.
The three largest groups in terms of number of members have $N_\mathrm{gal}=\{51,36,29\}$, $z=\{0.3742,0.7301,0.2204\}$, and $M_\mathrm{halo}=\{10^{14.97},10^{14.95},10^{13.62}\}$ $\mathrm{M}_\odot h^{-1}$.
Figure \ref{fig:D10_RAz} shows a summary view of the galaxy groups we find in DEVILS-D10.

%%%%%%%%%%%%%%%%%%%%%%%%%%%%%%%%%%%%%%%%%%%%%%%%%%
\subsection{Comparison with the zCOSMOS 20k Group Catalogue}\label{S5.1:lit_comp}

\begin{figure}
    \centering
    \includegraphics[width=\linewidth]{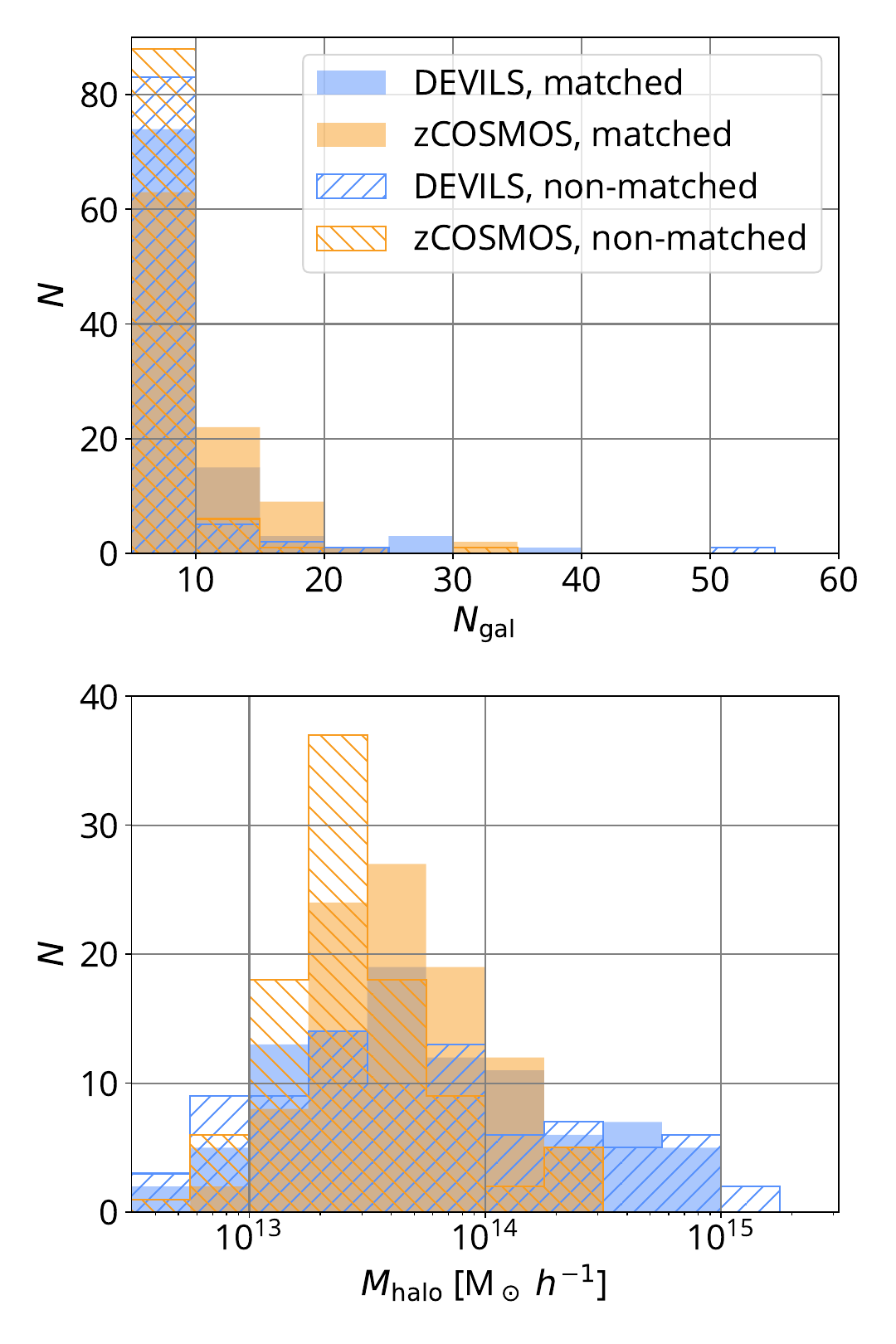}
    \caption{Comparison of the distributions of the number of members ($N_\mathrm{gal}$, left) and halo masses ($M_\mathrm{halo}$, right) of the groups with at least five members ($N_\mathrm{gal}\geq5$) from our DEVILS Group Catalogue and the zCOSMOS 20k Group Catalogue \citep{knobel2012}, divided into matched and non-matched groups.
    Galaxy groups that have been matched are shown with solid histograms (blue for DEVILS and orange for zCOSMOS), and those that have no close matches are shown with hatched histograms (same colours as before).}
    \label{fig:D10GC_vs_zCOSMOS_all}
\end{figure}

\begin{figure}
    \centering
    \includegraphics[width=\linewidth]{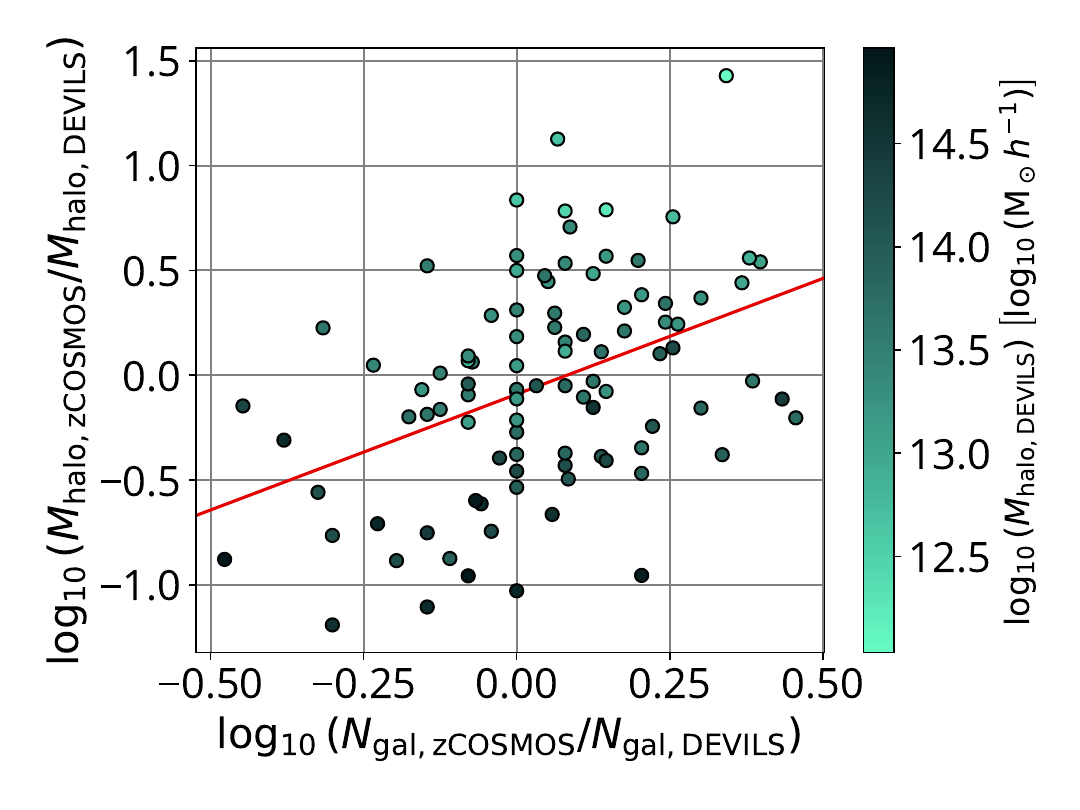}
    \caption{Comparison of the $N_\mathrm{gal}$ (left) and $M_\mathrm{halo}$ (right) for the matched galaxy groups between the zCOSMOS 20k Group Catalogue \citep{knobel2012} and our DEVILS-D10 Group Catalogue.
    Markers are coloured by the halo mass in DEVILS-D10, and the red line shows the best-fit linear relation.}
    \label{fig:D10GC_vs_zCOSMOS_matched}
\end{figure}

The DEVILS-D10 Group Catalogue we present here is the latest effort to provide a galaxy group catalogue in the COSMOS field, being preceded by several previous works \citep[e.g.,][]{george2011,knobel2012,sochting2012,toni2024,toni2025}.
Of these previous efforts, the zCOSMOS 20K Group catalogue \citep{knobel2012} is the most similar to our DEVILS-D10 Group Catalogue.
Briefly, the zCOSMOS 20 Group Catalogue was built using the zCOSMOS-bright redshift survey \citep{lilly2007,lilly2009}, which targeted $\sim20,000$ objects in the $15\leq i_\mathrm{ap}\leq22.5$ range with the VIsible MultiObject Spectrograph \citep[VIMOS;][]{LeFevre2003} on the Very Large Telescope.
Of these targets, $\sim48\%$ got secure spectroscopic redshifts, with the rest of the sample having photometric redshifts derived using a mix of 26 broad, medium, and narrow band filters covering from the optical to the mid-infrared.

The groups are then found using the FoF algorithm presented in \citet{knobel2009}, which is simpler compared to the \citetalias{robotham2011a} algorithm, with 5 free parameters\footnote{While the \citet{knobel2009} FoF algorithm proper has only 3 free parameters, \citet{knobel2009,knobel2012} use different values of these parameters to recover groups of different number of members, which can be counted as 2 extra free parameters ($N_\mathrm{gal,min}$ and $N_\mathrm{gal,max}$).}.
A second group finding method, employing a Voronoi–Delaunay algorithm \citep[e.g.,][]{marinoni2002,gerke2005}, was used to measure the robustness of the finding of each group \citep[\texttt{GRP2}, see sections 3.2 and 3.3 of][for further details]{knobel2012}.
The photometric redshifts are used to calculate the probability of galaxies belonging to spectroscopically-found groups, with the corrected memberships then used to estimate halo masses.

To compare both catalogues, we match the group galaxies with $N_\mathrm{gal}\geq5$ in each by position in the sky and redshift, for which we use the \texttt{match\_coordinates\_sky} function provided in \textsc{astropy} to find the closest match across catalogues.
We then define groups as matched if the projected separation is lower than 0.7 Mpc and if the difference in redshift is lower than 0.01 (i.e, $d_\mathrm{2D}<0.7$ Mpc $\bigwedge$ $|\Delta z|<0.01$), based on exploration of the distribution of both $d_\mathrm{2D}$ and $|\Delta z|$.
We find 97 groups that match this criteria, leaving 93 (96) groups in the DEVILS-D10 Group Catalogue (zCOSMOS 20k Group Catalogue) without a counterpart.
We note that we find that the corrected memberships using photometric redshifts are in poor agreement with those from DEVILS (being highly overestimated), with the pure spectroscopic memberships in better agreement, so we focus on the latter here.

We explore the differences between matched and non-matched groups in Figure \ref{fig:D10GC_vs_zCOSMOS_all}, which we combine with statistical comparisons using the 2-sample Anderson-Darling test implementation in \textsc{scipy}\footnote{Using the permutation-based calculation suggested for cases with low sample sizes. \url{https://docs.scipy.org/doc/scipy/reference/generated/scipy.stats.anderson_ksamp.html}.} (\texttt{stats.anderson\_ksamp}).
Both catalogues show the same trend of the non-matched groups being biased towards lower $N_\mathrm{gal}$ compared to the matched groups ($p^{}_\mathrm{AD}\leq10^{-4}$ between matched and non-matched $N_\mathrm{gal}$ distributions), as we expect due to small groups being more susceptible to contamination from interlopers in the hybrid spectroscopic-photometric redshift sample used by \citet{knobel2012}.
While zCOSMOS shows a statistical difference between the matched and non-matched groups in terms of halo mass and DEVILS-D10 does not ($p^{}_\mathrm{AD}\leq10^{-4}$ and $p^{}_\mathrm{AD}=0.1398$, respectively), we note that the difference is relatively minor, with a difference in median $M_\mathrm{halo}$ of $\sim0.15$ dex.
For zCOSMOS, there are no clear differences between the matched and non-matched groups in the distributions of RA, declination, $z$, and \texttt{GRP2} ($p^{}_\mathrm{AD}>0.05$ for all four properties), while for DEVILS-D10 there is a statistically-meaningful difference between the RA of matched and non-matched groups ($p^{}_\mathrm{AD}=0.0311$), with the latter being more centrally-concentrated around $150^\circ$ than the former.
Combined, these results suggest that the differences between zCOSMOS and DEVILS stem from the significantly improved spectroscopic redshift completeness of the latter and not from any particular group property.

Comparing the properties of the groups that we have matched across DEVILS-D10 and zCOSMOS, the Spearman rank correlation coefficient\footnote{Calculated with the \textsc{scipy} implementation (\texttt{stats.spearmanr}), following the suggested implementation for low sample sizes \url{https://docs.scipy.org/doc/scipy/reference/generated/scipy.stats.spearmanr.html}.} between catalogues for both $N_\mathrm{gal}$ and $M_\mathrm{halo}$ have values of $\rho^{}_\mathrm{Spearman}\sim0.5$ (and $p^{}_\mathrm{Spearman}\ll0.05$), indicating a mild correlation between the relative rankings of groups in both catalogues.
We also find little relative bias for both $N_\mathrm{gal}$ and $M_\mathrm{halo}$, with median offsets of $\sim0.05$ dex (DEVILS-D10 having slightly fewer galaxies but slightly larger halo masses), but with a significant scatter of $\sim0.5$ dex.
Both quantities show similar overall trends, being over/under-estimated in zCOSMOS relative to DEVILS for small/large groups.
These two trends are connected, as shown in Figure \ref{fig:D10GC_vs_zCOSMOS_matched}, with a Pearson correlation coefficient between $\log_{10}(N_\mathrm{g,zCOSMOS}/N_\mathrm{g,DEVILS})$ and $\log_{10}(M_\mathrm{halo,zCOSMOS}/M_\mathrm{halo,DEVILS})$ of $r^{}_\mathrm{P}=0.4160$ indicating that groups with too many members in zCOSMOS have their masses overestimated.

\begin{figure}
    \centering
    \includegraphics[width=\linewidth]{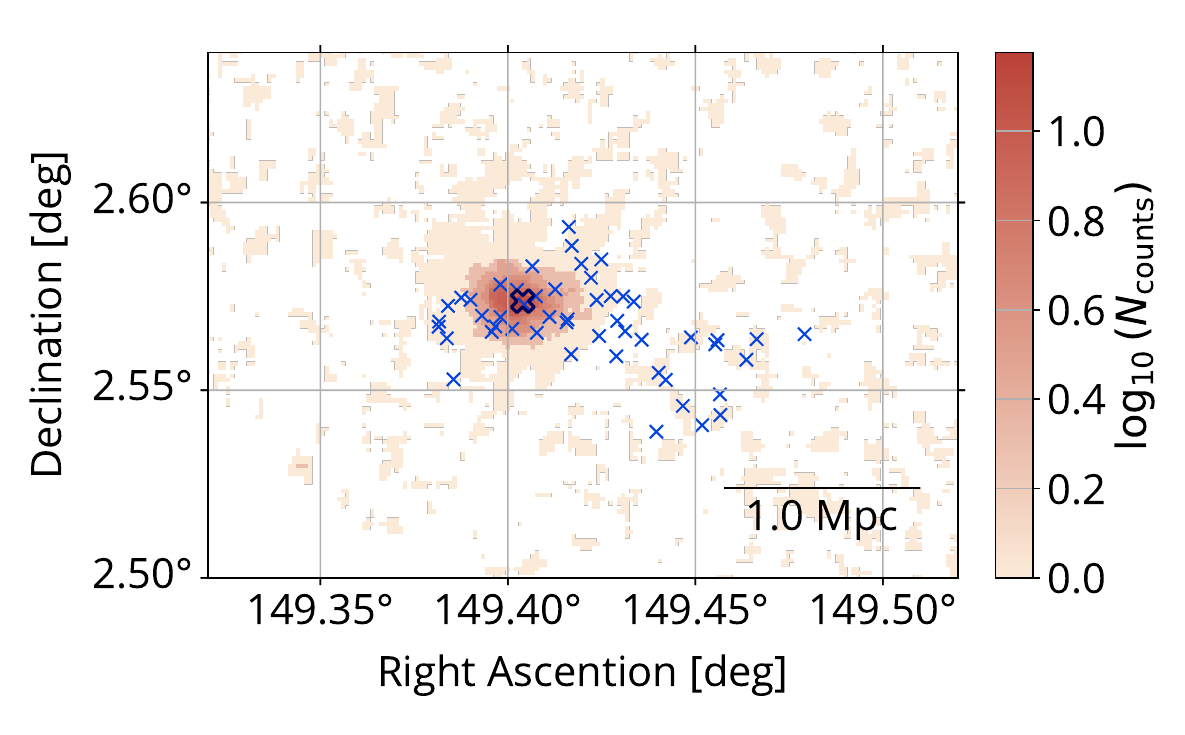}
    \caption{X-ray emission from the \textit{Chandra}-COSMOS Legacy survey \citep{civano2016} in the location of the largest cluster in the DEVILS-D10 Group Catalogue.
    The image has been smoothed using a 5x5 pixel median filter, to highlight the emission from extended sources.
    The large, open cross marker indicates the position of the central galaxy of the cluster, with the smaller crosses indicating the position of the satellites.}
    \label{fig:D10_Chandra}
\end{figure}

We note that the largest cluster (both by member count and mass) in our DEVILS-D10 Group Catalogue is not matched to any group in the zCOSMOS 20k Group Catalogue.
This cluster lies near the eastern edge of the DEVILS-D10 field, where the zCOSMOS catalogue is notably sparser (by a factor of $\sim6$), with no galaxies within $\Delta z<0.1$ of the cluster.
Given our estimated mass of $M_\mathrm{halo}\sim10^{15}$ $\mathrm{M}_\odot h^{-1}$, we expect this cluster to be detectable through X-ray emission, which we also expect to be well-aligned with the central galaxy of the cluster based on the findings by \citet{popesso2024}.
We explore the archival data from the \textit{Chandra}-COSMOS Legacy Survey \citep{civano2016} in the location of the cluster, were we find clear X-ray emission centred on the cluster central, validating this particular group in the DEVILS-D10 Group Catalogue.

%%%%%%%%%%%%%%%%%%%%%%%%%%%%%%%%%%%%%%%%%%%%%%%%%%
\section{AGN in isolated and satellite galaxies}\label{S6:agn_env}

As a showcase of the science that will be enabled by this group catalogue, we explore the impact that environment plays on the presence of active galactic nuclei (AGN) in galaxies.
There is no clear consensus in the literature on the impact of environment on AGN, with some works pointing towards reduced fraction of AGN in satellite galaxies \citep[e.g.,][]{hou2024} while others report no significant difference \citep{man2019,wethers2022}, even though there is evidence of environmental processes enhanced AGN activity \citep[e.g.,]{poggianti2017,peluso2022}.
\citet{thorne2022} presented the AGN properties derived using SED fitting for DEVILS-D10 galaxies, with \citet{dsilva2023a} finding a strong coevolution of the cosmic densities of both SFR and AGN, and \citet{davies2025b} finding a strong connection between the presence of AGN and the declining star formation histories in massive galaxies.

\begin{figure}
    \centering
    \includegraphics[width=\linewidth]{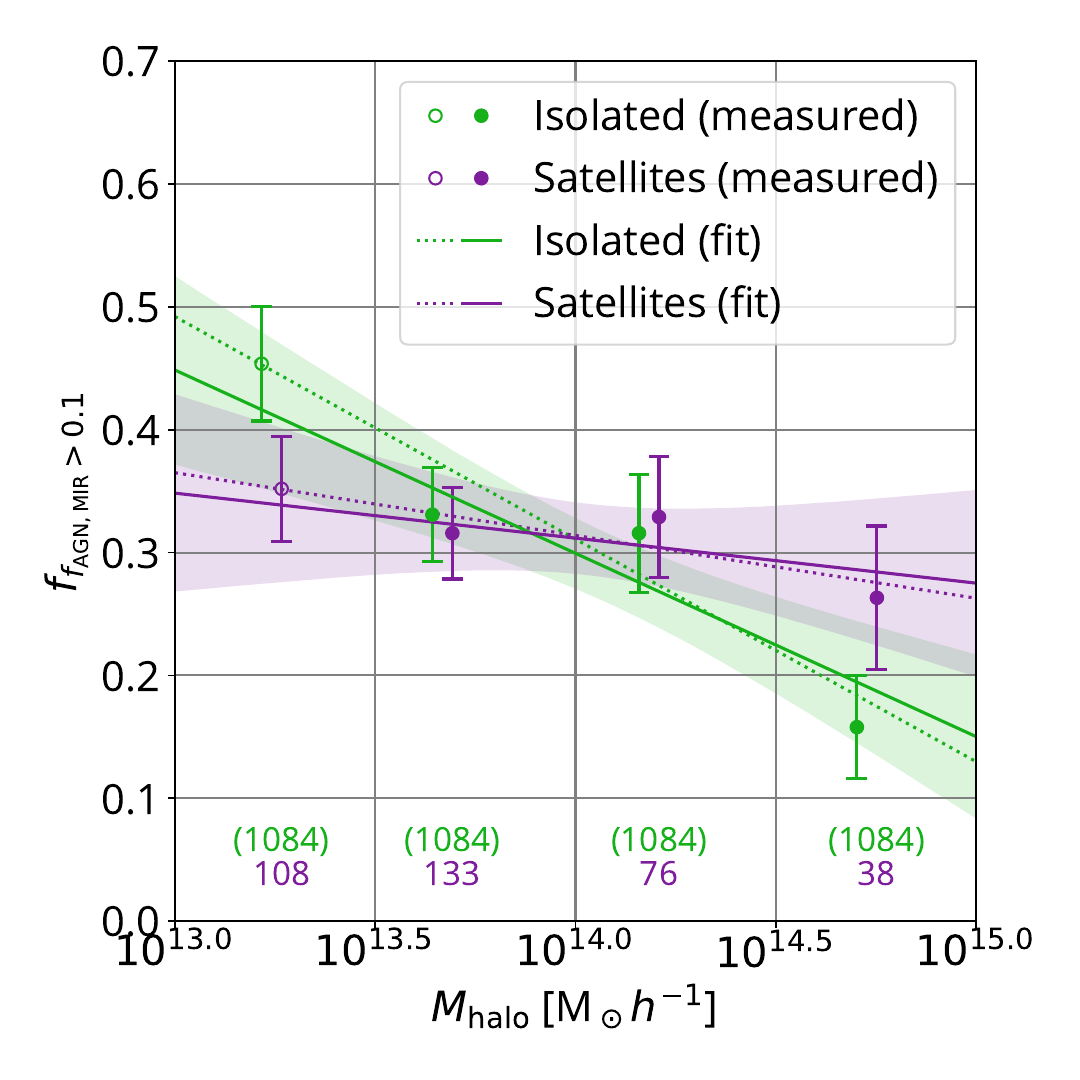}
    \caption{Fraction of the galaxies with $10^{11}<M_\star/\mathrm{M}_\odot<10^{11}$ at $0.3<z<0.5$ with a significant AGN contribution in the rest-frame mid-IR \citep[$f_\mathrm{AGN,MIR}>0.1$ as defined by][]{thorne2022}, as a function of halo mass.
    Galaxies are divided between isolated (green markers and lines) and satellite galaxies from groups with $N_\mathrm{gal}\geq3$ (purple markers and lines), and into four halo mass bins.
    The markers indicate the measured AGN fractions in each halo mass bin, with 68\% confidence intervals calculated following \citet{cameron2011}.
    The markers for isolated galaxies are offset to the left for visualisation purposes, and the open markers indicate the $M_\mathrm{halo}$-incomplete bins.
    The upper numbers indicate the parent sample of isolated galaxies, and the lower numbers the satellite and $M_\star$-matched isolated galaxies in each bin.
    The solid (dotted) lines show the best linear fits in the $M_\mathrm{halo}$-complete range (for all bins), with the 68\% uncertainty in shaded areas.}
    \label{fig:DEVILS-D10_fAGN}
\end{figure}

Here we compare the AGN properties of isolated and satellite galaxies following the same stellar and redshift selection used by \citet{davies2025c}, selecting galaxies with $10^{10}<M_\star/\mathrm{M}_\odot<10^{11}$ in the $0.3<z<0.5$ range, which ensures a $M_\star$-complete sample \citep[see figure 1 of ][]{davies2025c}.
Since the \citetalias{robotham2011a} group finder does not estimate halo masses for isolated galaxies, we create a comparison sample by $M_\star$-matching each satellite to an isolated galaxy in our chosen redshift and stellar mass range.  
We include satellite galaxies only from groups with $N_\mathrm{g}\geq3$.
We note that, with these selections, we are $M_\mathrm{halo}$-complete only for $M_\mathrm{halo}\gtrsim\mstar{13.5}h^{-1}$.

\begin{figure}
    \centering
    \includegraphics[width=\linewidth]{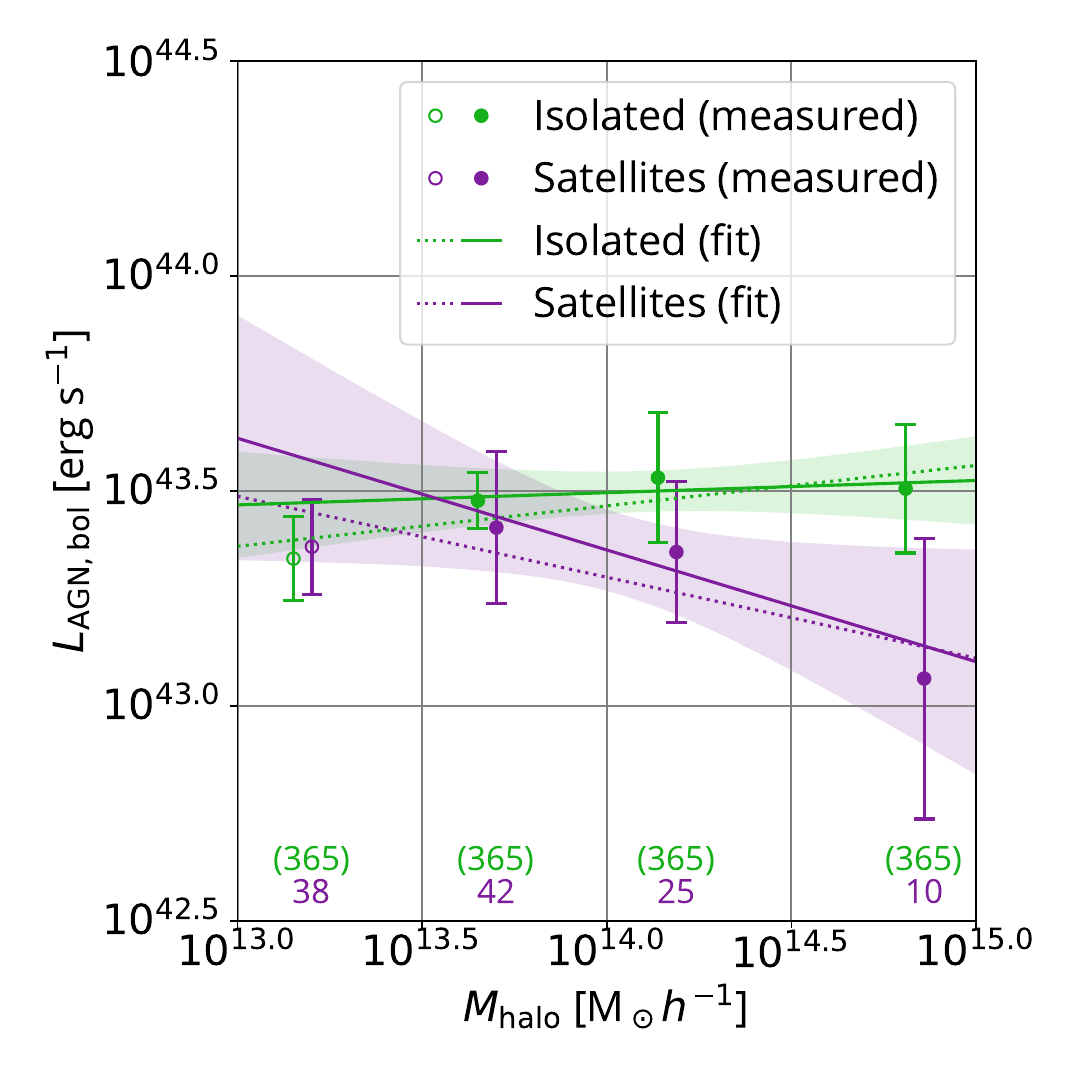}
    \caption{Median AGN bolometric luminosity ($L_\mathrm{AGN}$) for isolated and satellite galaxies with a significant AGN contribution in the mid-IR, as a function of halo mass.
    Markers, lines, colours, and numbers as in Figure \ref{fig:DEVILS-D10_fAGN}.
    The 68\% confidence intervals for the medians are calculated using the bootstrapping function implemented in \textsc{scipy}.}
    \label{fig:DEVILS-D10_Lbol}
\end{figure}

We first explore the fraction of galaxies with significant AGN, following \citet{thorne2022} by first defining the fraction of the mid-IR flux that comes from the AGN ($f^{}_\mathrm{AGN,MIR}=F_\mathrm{MIR,AGN}/F_\mathrm{MIR,tot}$), and then calculating the fraction of galaxies where AGN contribute at least 10\% of the flux ($f^{}_\mathrm{AGN,MIR}>0.1$).
Figure \ref{fig:DEVILS-D10_fAGN} shows the resulting AGN fractions as a function of halo mass.
When compared to isolated galaxies of similar stellar mass, we find that satellite galaxies in large groups/small clusters ($10^{13.5}< M_\mathrm{halo}/(\mathrm{M}_\odot h^{-1})<10^{14.5}$) show similar AGN fractions.
In contrast, satellite galaxies in massive clusters ($M_\mathrm{{halo}}>\mstar{14.5}h^{-1}$) exhibit an increase in the AGN fraction compared to isolated galaxies of comparable mass (from $\sim0.15$ to $\sim0.25$), while low-mass groups ($M_\mathrm{{halo}}<\mstar{13.5}h^{-1}$) show the reserve with a diminished AGN fraction in satellites compared to isolated galaxies (from $\sim0.45$ to $\sim0.35$).
Figure \ref{fig:DEVILS-D10_Lbol} shows the median AGN bolometric luminosities for galaxies with $f^{}_\mathrm{AGN,MIR}>0$ as a function of halo mass.
We find that satellites in clusters ($M_\mathrm{{halo}}>\mstar{14}h^{-1}$) exhibit a decreased median luminosity of AGN relative to isolated galaxies, by up to $\sim0.5$ dex, while in groups ($M_\mathrm{{halo}}<\mstar{14}h^{-1}$) the median luminosities appear consistent with no environmental impact.

Combining our measurements presented in Figures \ref{fig:DEVILS-D10_fAGN} and \ref{fig:DEVILS-D10_Lbol}, we find evidence that environment has a measurable impact on the presence and strength of AGN in satellite galaxies, modulated by the halo mass of the group/cluster.
While we are cautious about interpreting our measurements for small groups, we find that large groups do not significantly impact the AGN on satellite galaxies.
Satellites in clusters, in particular those in massive ones, diverge from the trends seen in isolated galaxies, with AGN becoming more common but dimmer.
The increased AGN fraction in clusters appears consistent with previous works showing increased AGN activity in jellyfish galaxies in $z\sim0$ clusters \citep[e.g.,][]{poggianti2017,peluso2022}, suggesting that the strong ram pressure generated in massive clusters can help funnel gas into the SMBH of satellite galaxies, though how this is connected to the decrease in luminosity we find is not clear.
We leave a more detailed exploration for future work, including the addition of the X-ray and radio data available in COSMOS to further explore the presence of AGN.

%%%%%%%%%%%%%%%%%%%%%%%%%%%%%%%%%%%%%%%%%%%%%%%%%%
% Conclusions
%%%%%%%%%%%%%%%%%%%%%%%%%%%%%%%%%%%%%%%%%%%%%%%%%%
\section{Conclusions}\label{S5:conclusion}

In this work, we have presented the first version of the galaxy group catalogue for the DEVILS D10-COSMOS field, covering a redshift range of $0\lesssim z\lesssim1.0$ (the last $\sim7$ Gyr) with a 90\% redshift completeness down to $Y=21.2$.
To generate this catalogue, we have used a lightly updated version of the same group finder used for the GAMA survey, presented in \citet{robotham2011a}, and calibrated it using custom-made DEVILS-like synthetic lightcones using the latest version of the \shark\ semi-analytic model \citep{lagos2024,chandro2025}.
The D10 galaxy group catalogue contains $15222$ galaxies, with $5921$ being members of one of the $2060$ galaxy groups in D10, $189$ of which have at least $5$ members.
We use the group catalogue to explore the impact of environment on the occurrence and strength of AGN, finding that $0.3<z<0.5$ satellites in groups show similar AGN fractions and luminosities as those of similar isolated galaxies, while satellites in clusters exhibit increased AGN fractions but with decreased luminosities.
The DEVILS-D10 Group Catalogue will be made publicly available as part of DEVILS Data Release 1 (Davies et al., in preparation).

%%%%%%%%%%%%%%%%%%%%%%%%%%%%%%%%%%%%%%%%%%%%%%%%%%
% Acknowledgements
%%%%%%%%%%%%%%%%%%%%%%%%%%%%%%%%%%%%%%%%%%%%%%%%%%
\begin{acknowledgments}
MB is funded by McMaster University through the William and Caroline Herschel Fellowship.
LJMD and ASGR acknowledge support from the Australian Research Council's Future Fellowship and Discovery Project schemes (FT200100055, FT200100375 and DP250104611).
MS acknowledges support by the State Research Agency of the Spanish Ministry of Science and Innovation under the grants `Galaxy Evolution with Artificial Intelligence' (PGC2018-100852-A-I00) and `BASALT' (PID2021-126838NB-I00) and the Polish National Agency for Academic Exchange (Bekker grant BPN/BEK/2021/1/00298/DEC/1).
TSL acknowledges support from the Australian Research Council (ARC) Laureate Fellowship scheme (FL220100191).
This work was supported by resources provided by the Pawsey Supercomputing Research Centre’s Setonix Supercomputer (\url{https://doi.org/10.48569/18sb-8s43}) and Acacia Object Storage (\url{https://doi.org/10.48569/nfe9-a426}), with funding from the Australian Government and the Government of Western Australia.
This work was partially supported by the European Union's Horizon 2020 Research and Innovation program under the Maria Sklodowska-Curie grant agreement (No. 754510).
The X-ray data used to confirm the largest cluster in DEVILS-D10 was obtained from the Chandra Data Archive \dataset[(Chandra ObsId 15259)]{https://doi.org/10.25574/15259}.
\end{acknowledgments}


\vspace{5mm}
\facilities{AAT (AAOmega/2dF), VISTA (VIRCAM)}


\software{\textsc{python} v3.11 (\url{https://www.python.org}),
          \textsc{astropy} v7.1 \citep{astropyI,astropyII,astropyIII},
          \textsc{h5py} v3.14 \citep{h5py},
          \textsc{jupyter} v1.0 \citep{jupyter},
          \textsc{matplotlib} v3.10 \citep{matplotlib},
          \textsc{numpy} v2.3 \citep{numpy},
          \textsc{pandas} v2.3 \citep{pandas},
          \textsc{scicm} v1.0 \citep{scicm},
          \textsc{scipy} v1.16 \citep{scipy},
          \textsc{splotch} v0.6 \citep{splotch},
          \textsc{stingray} v0.43 \citep{stingray_043},
          \textsc{R} v4.3 \citep{R},
          \textsc{celestial} v1.4 \citep{celestial},
          \textsc{fof} v1.2 \citep{fof_122},
          \textsc{highlander} v0.2 \citep{highlander_0210},
          \textsc{prospect} v1.6 \citep{robotham2020},
          \textsc{viperfish} v0.5 \citep{viperfish_053}}

%%%%%%%%%%%%%%%%%%%%%%%%%%%%%%%%%%%%%%%%%%%%%%%%%%
% Appendices
%%%%%%%%%%%%%%%%%%%%%%%%%%%%%%%%%%%%%%%%%%%%%%%%%%
%\appendix

%\section{Example groups}\label{examples}

%This is a totally non-biased selection of groups to show how nice they are.

%%%%%%%%%%%%%%%%%%%%%%%%%%%%%%%%%%%%%%%%%%%%%%%%%%

\bibliography{papers}{}
\bibliographystyle{aasjournal}



\end{document}